\documentclass[a4paper,useAMS,usenatbib]{mnras}
\pdfoutput=1
\usepackage{newtxtext,newtxmath}
\usepackage[T1]{fontenc}
\usepackage{ae,aecompl}
\usepackage{graphicx}	
\usepackage{amsmath}	
\usepackage{amssymb}	

\title[Hercules-Aquila and Virgo Clouds with Gaia DR2]{Common origin
  for Hercules-Aquila and Virgo Clouds in Gaia DR2}

\author[Iulia T. Simion et al]{Iulia T. Simion$^{1}\thanks{E-mail:isimion@shao.ac.cn}$, Vasily Belokurov$^{2,3}$ and  Sergey E. Koposov$^{4,2}$\\
  $^{1}$Key Laboratory for Research in Galaxies and Cosmology, Shanghai Astronomical Observatory, 80 Nandan Road, Shanghai 200030, China\\
  $^{2}$Institute of Astronomy, Madingley Rd, Cambridge, CB3 0HA\\
  $^{3}$Center for Computational Astrophysics, Flatiron Institute, 162 5th Avenue, New York, NY 10010, USA\\
  $^4$McWilliams Center for Cosmology, Department of Physics, Carnegie Mellon University, 5000 Forbes Avenue, Pittsburgh, PA, 15213, USA}

\date{Accepted XXX. Received YYY; in original form ZZZ}

\pubyear{2018}

\begin{document}
\label{firstpage}
\pagerange{\pageref{firstpage}--\pageref{lastpage}}
\maketitle

\begin{abstract}
We use a sample of $\sim$350 RR Lyrae stars with radial velocities and
Gaia DR2 proper motions to study orbital properties of the
Hercules-Aquila Cloud (HAC) and Virgo Over-density (VOD). We
demonstrate that both structures are dominated by stars on highly
eccentric orbits, with peri-centres around $\sim1$ kpc and apo-centres
between 15 and 25 kpc from the Galactic centre. Given that the stars
in the HAC and the VOD occupy very similar regions in the space
spanned by integrals of motion, we conclude that these diffuse debris
clouds are part of the same accretion event. More precisely, these
inner halo sub-structures likely represent two complementary
not-fully-mixed portions of an ancient massive merger, also known as
the ``sausage'' event.
\end{abstract}

\begin{keywords}
Galaxy: structure -- Galaxy : formation -- galaxies: individual: Milky
Way.
\end{keywords}



\section{Introduction}
How do you hide the evidence for a massive impact event that caused
the extinction of most of the dinosaurs as well as 75\% of all species
on Earth? You bury it deep under the sea, covered with a layer of
sediment taller than the Empire State Building
\citep[][]{Hildebrand1991}. Without the discovery of the giant
Chicxulub crater, the meteorite impact hypothesis would remain a neat
theory supported by striking but indirect evidence. A hypothesis of an
ancient dramatic collision between the Milky Way and an unidentified
massive dwarf galaxy was put forward by \citet{Deason2013} to explain
a particular feature in the overall stellar halo density profile
\citep[][]{Wa09,Sesar2011,De11}. Most recently, through a study of a
portion of the nearby stellar halo, \citet{Belokurov2018} demonstrated
that the great impactor must have collided with the young Milky Way on
a nearly radial orbit, thus swamping the inner stellar halo with
metal-rich material with orbital anisotropy \citep[see][]{Binney2008}
close to unity. Merger events like this tend to leave behind a battery
of debris clouds and shells \citep[see
  e.g.][]{Johnston2008,Amorisco2015,Hendel2015}, which - akin to the
peak rings of impact craters \citep[see e.g.][]{Morgan2016} - if
discovered could help to reconstruct the collision as well as pin down
the properties of the progenitor \citep[e.g][]{Sanderson2013,Johnston2016}.

Before the Data Release 2 \citep[][]{Brown2018} of the ESA's Gaia
mission \citep[][]{Prusti2016}, five large and diffuse cloud-like
structures had been discovered in the Galaxy's halo. These include:
the Virgo Over-Density
\citep[VOD,][]{Vivas2001,Newberg2002,Duffau2006,Juric2008,Bonaca2012},
the Hercules-Aquila Cloud \citep[HAC,][]{Be07,Simion2014}, the
Trinagulum-Andromeda structure
\citep[Tri-And,][]{Rocha2004,Majewski2004,Deason2014}, the Pisces
Over-density \citep[][]{Sesar2007,Wa09,Nie2015} and the
Eridanus-Phoenix over-density \citep[Eri-Pho,][]{Li2016}. According to
the most recent investigations, Tri-And likely comprises of Galactic
disc stars kicked out of the plane in a recent interaction with a
dwarf galaxy, probably the Sagittarius dSph
\citep[e.g.][]{Pr15,Bergemann2018,Hayes2018}. Of the remaining four,
the Pisces overdensity clearly stands out as it reaches much larger
Galacto-centric distances. On the other hand, the VOD, HAC and Eri-Pho
structures occupy a very similar range of distances, between 10 and 20
kpc from the Galactic center. This led \citet{Li2016} to suggest that
these three Clouds could all be part of one merger event, a galaxy
accreted onto the Milky Way on a polar orbit \citep[see also][]{Juric2008}.

As demonstrated by the recent re-interpretation of the Monoceros Ring
(and the associated sub-structures) and the Tri-And, deciphering the
nature of halo over-densities is often impossible without either
high-resolution spectroscopy \citep[e.g.][]{Bergemann2018} or accurate
astrometry \citep[e.g.][]{deBoer2018,Deason2018}. In this Letter, we
look for clues to the formation of the Hercules-Aquila and Virgo
Clouds using proper motions provided as part of the Gaia DR2. At our
disposal are highly pure samples of members of each Cloud, namely the
RR Lyrae stars that i) are co-spatial with HAC and VOD in 3-D and ii)
that have their line-of-sight velocities measured. By complementing
the publicly available 4-D data with the GDR2 proper motions, we build
a large tracer set with complete 6-D phase space information and study
the make-up of each structure using the orbital properties of the
constituent stars.
\section{Data and analysis}
\begin{figure}
	\includegraphics[scale=0.518]{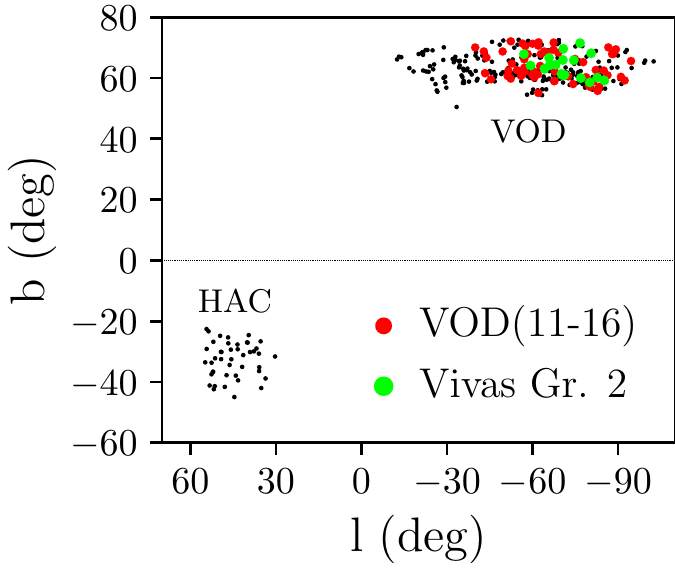}
	\includegraphics[scale=0.518]{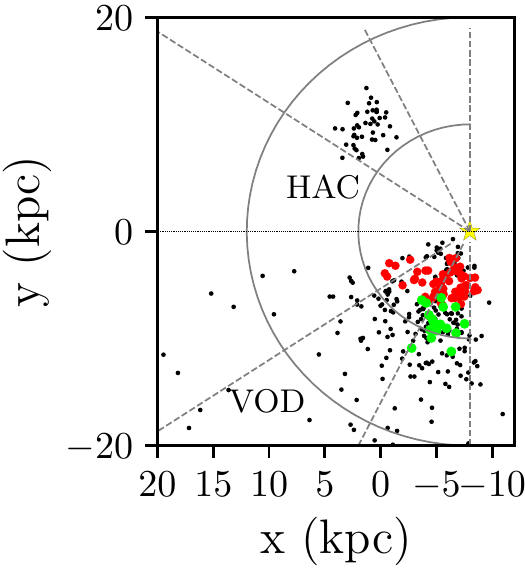}
	\includegraphics[scale=0.518]{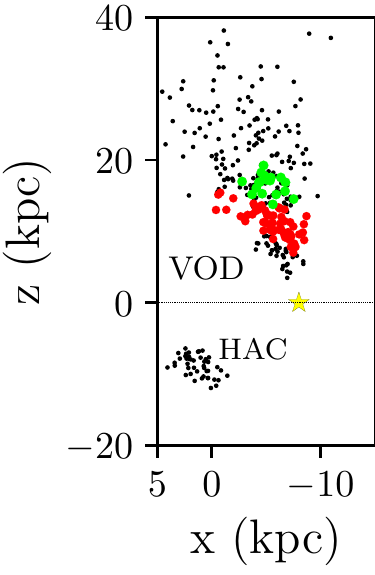}
	\vspace{-0.55cm}
    \caption{Spatial distribution of $\sim$350 RR Lyrae with 6-D phase
      space information in the HAC and VOD fields, in Galactic
      coordinates (left panel) and in the x-y (middle) and x-z (right)
      planes. In green we mark a significant kinematical group
      identified by \citet{Vivas2016} (Group~2 in their Table 5) and
      in red a sub-sample of the VOD RRL with galactocentric distances
      similar to the HAC sample, 11 $<\mathrm{r_{GC}}$/kpc$<$
      16. The semi-circles are centred on the Sun's position and have
      a radius of 10 and 20 kpc. The Sun (yellow star) is located at
      (x$_{\odot}$, y$_{\odot}$, z$_{\odot})= $ (-8,0,0) kpc.}
    \label{fig:lb}
\end{figure}
\subsection{4-D RR Lyrae data}
The Hercules-Aquila and Virgo Clouds are diffuse stellar
over-densities in the inner stellar halo, located on the opposite
sides of the Galaxy (see Figure~\ref{fig:lb}). At high latitudes,
these are detected as peaks in RR Lyrae number counts - curiously - at
similar heliocentric distances, i.e. $\sim$17 kpc (HAC:
\citealt{Wa09,Simion2014}) and $\sim$19 kpc (VOD: \citealt{Vivas2006,
  Duffau2014, Vivas2016}). Note that other tracers (e.g. BHBs, MSTO
and K and M giants) have also been used to pin down the morphology of
the Clouds \citep[see
  e.g.][]{Be07,Juric2008,Sharma2010,Bonaca2012,Conroy2018}. The RR
Lyrae, however offer the clearest view of these halo sub-structures
thanks to their associated accurate distances and minuscule Galactic
foreground contamination. Therefore, in this work, we have focused on
the two recently published samples of RR Lyrae towards the Clouds,
where each star has a well-measured line-of-sight
velocity. \citet{Simion2018} provide a table of 46 RRL with radial
velocity measurements (45 observed at the Michigan-Dartmouth-MIT
Observatory and 1 from SDSS) with heliocentric distances between 15
and 18 kpc. \cite{Vivas2016} compiled a catalog of 412 RRL in the
region of the sky covered by the VOD with distances between 4 and 75
kpc from the Sun with radial velocity measurements of stars from La
Silla-QUEST, QUEST, CRTS and LINEAR.

\subsection{From 4-D to 6-D. Velocity distributions}
By cross-matching to the GDR2 data with an aperture of 2$''$, we have
found Gaia counterparts to 44 HAC stars and 411 VOD. From the VOD
sample, we remove 112 stars likely belonging to the Sgr stream as
identified by \citep[][, their Group 1]{Vivas2016}. The spatial
distribution of the remaining stars (44 from \citealt{Simion2018} and
299 from \citealt{Vivas2016}) with full 6-D phase space measurements
is given in Figure~\ref{fig:lb}, in Galactic coordinates in the left
panel and in the x-y (x-z) Galactic plane in the middle (right)
panel. We adopt a left-handed Galactic Cartesian coordinates with the
Sun located at (x$_{\odot}$,y$_{\odot}$, z$_{\odot}$) = (-8,0,0) kpc,
the x-axis positive in the direction of the Galactic center, y-axis
oriented along the Galactic rotation and the z-axis directed towards
the north Galactic pole. While \citealt{Vivas2016} identify 6
significant kinematical groups in the VOD region (their table 5), only
Group 1 (Sagittarius stream) and 2 (likely members of the VOD, with
$\mathrm{<v_{GSR}>}= 135$ km/s) contain more than 10 stars. We mark
Group 2 stars with green circles in Figure~\ref{fig:lb}. We also mark with
red circles the location of a group of stars selected to have
galactocentric distances similar to the HAC sample, i.e.
$11\mathrm{<r_{GC}}/$kpc$<16$ to facilitate a fair comparison of their
velocities and orbits.

\begin{figure}
	\includegraphics[scale=0.55]{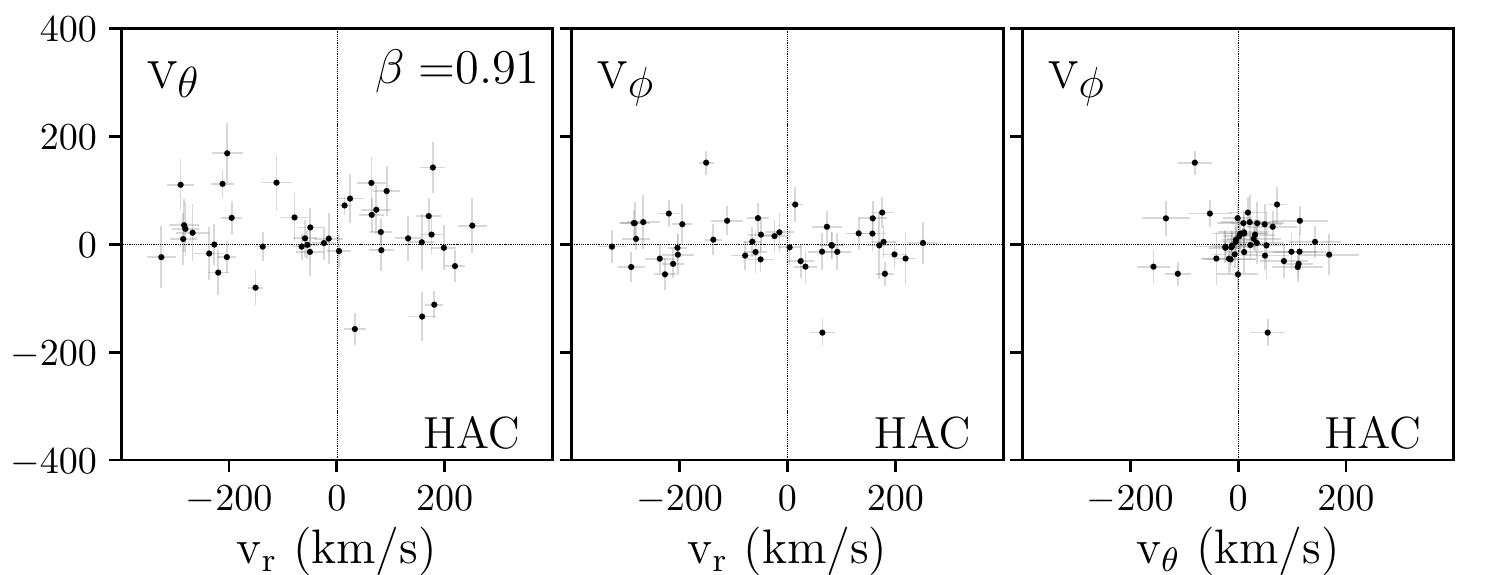}
  \includegraphics[scale=0.55]{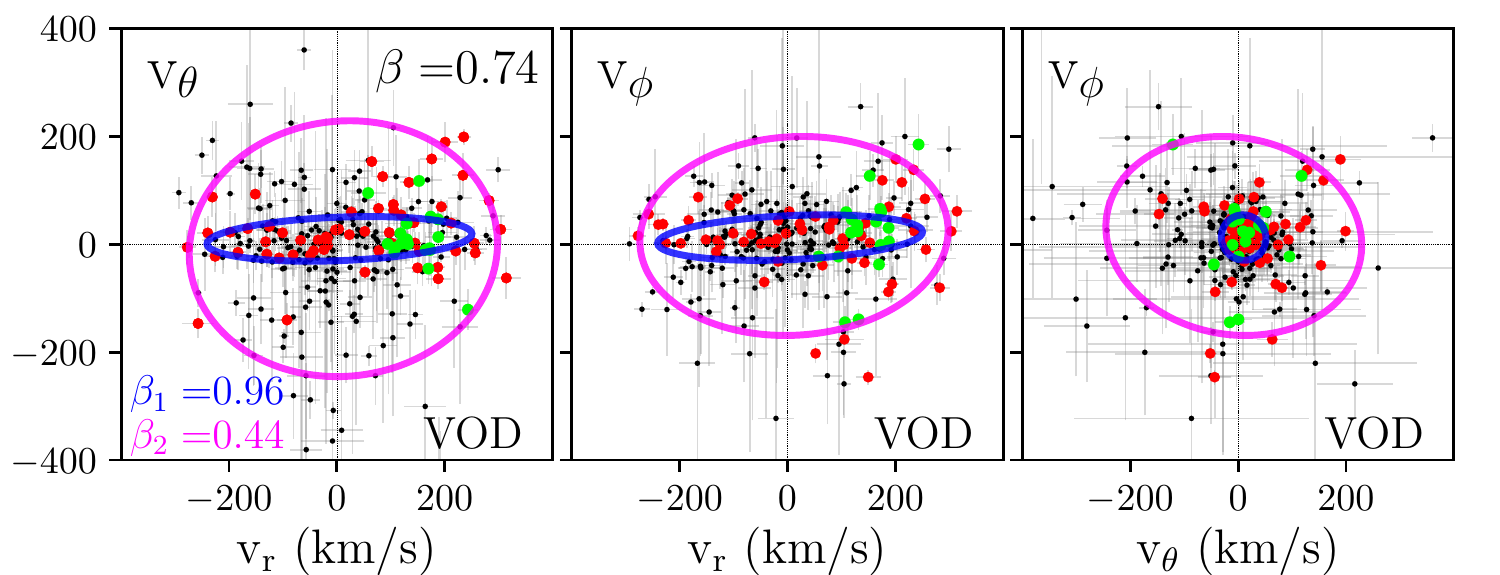}
   \vspace{-0.45cm}
    \caption{Velocity distribution in spherical polar coordinates
      ($\mathrm{v_{r}}$, $\mathrm{v_{\theta}}$, $\mathrm{v_{\phi}}$
      are the radial, azimuthal and polar components respectively, in
      km/s) of the HAC (top panels) and VOD (bottom panels) RR Lyrae.
      The label for the y-axis is reported in the upper-left corner of
      each panel. The uncertainties on the individual velocities (gray
      bars) were obtained propagating the errors on the measurements
      using Monte-Carlo methods. Both fields have radially biased
      orbits with average anisotropy parameters $\beta= 0.91 \pm 0.03$ (HAC)
      and $\beta= 0.74 \pm 0.04$ (VOD). A mixture of two multivariate
      Gaussians was fitted to the VOD data, according to which
      $\sim$62\% of the stars belong to the $\beta_{1}=
      0.96^{+0.02}_{-0.44}$ (blue) component and $\sim$38\% to the
      $\beta_{2}=0.44^{+0.45}_{-0.20}$ (magenta) component. The Vivas
      Group~2 and VOD (11$<$z/kpc$<$16) sub-samples are also shown in lime and
      red, as in Fig.\ref{fig:lb}. Vivas Group~2 clusters lie at $v_{r} =
      135$ km/s, as expected, while the VOD(11-16) group largely
      belongs to the high-anisotropy component (in blue).}
    \label{fig:vel}
\end{figure}

Figure~\ref{fig:vel} shows the Galacto-centric spherical polar
components of the velocities (radial $v_{r}$, azimuthal $v_{\theta}$
and polar $v_{\phi}$) of stars in the HAC (top row) and VOD (bottom
row) fields. To compute $v_{r}$, $v_{\theta}$ and $v_{\phi}$ we have
used $\texttt{astropy}$ \citep{astropy} with the default values for
the Sun's motion. In the Figure, Group 2 stars (green) can be seen
clustering at $v_{r} = 135$ km/s (by design), while the stars at
intermediate $\mathrm{r_{GC}}$ (red) seem to have a velocity
distribution very similar to those in the HAC, shown in the top
row. To estimate the uncertainty in each velocity component we
propagate the measured proper motion and line-of-sight velocity errors
using Monte-Carlo re-sampling, where we take into account the
covariances between the measurements of the right ascension and
declination components of proper motion as provided in GDR2. We use
the standard deviations of the resulting \{$v_{r}$, $v_{\theta}$,
$v_{\phi}$\} distributions as an estimate of the velocity
uncertainties; these are shown in Fig.~\ref{fig:vel}.
%
%

As evident from the Figure, the velocity distributions are highly
anisotropic, with the dispersion in the radial component dominating
the tangential ones, especially in the HAC data. To describe the shape
of the velocity distributions, we model each stellar sample as a
single-component multivariate Gaussian using the Extreme
Deconvolution \citep{ED} method as implemented in $\mathrm{astroML}$
\citep{astroML} package. The resulting parameters and the associated
uncertainties of the velocity ellipsoids are taken to be the median
and the standard deviation values of 500 bootstrap resampling trials.
The velocity ellipsoid shape can be summarized using the anisotropy
parameter $\beta=1-(\sigma^2_{\theta}+\sigma^2_{\phi})/2\sigma^2_r$
\citep[see][]{Binney2008}. We find the HAC stars have radially biased
orbits with $\beta = 0.91 \pm 0.03$ while for the whole of the VOD
sample, $\beta = 0.74 \pm 0.04$. Note, however, that the two samples
span very different ranges in Galactic $l,b$ and distances. According
to \citet{Belokurov2018}, the inner stellar halo can be viewed as a
mixture of two debris components with distinct
properties. Accordingly, we fit a model with two multivariate
Gaussians to the VOD velocity data using Extreme Deconvolution. 
With log-likelihood of $\log L = -5119$, the
two-component model is clearly preferred compared to the
single-component one with $\log L = -5384$. The VOD sample appears to
be composed of roughly two thirds of stars with highly anisotropic
velocity distribution $\beta_{1}=
0.96^{+0.02}_{-0.44}$ (marked in blue in Fig. \ref{fig:vel}) and a
third with more isotropic velocities $\beta_{2}=0.44^{+0.45}_{-0.20}$ (magenta) in good
agreement with the results for the local halo presented in
\citet{Belokurov2018}.

\begin{figure}
	        \includegraphics[scale=0.55]{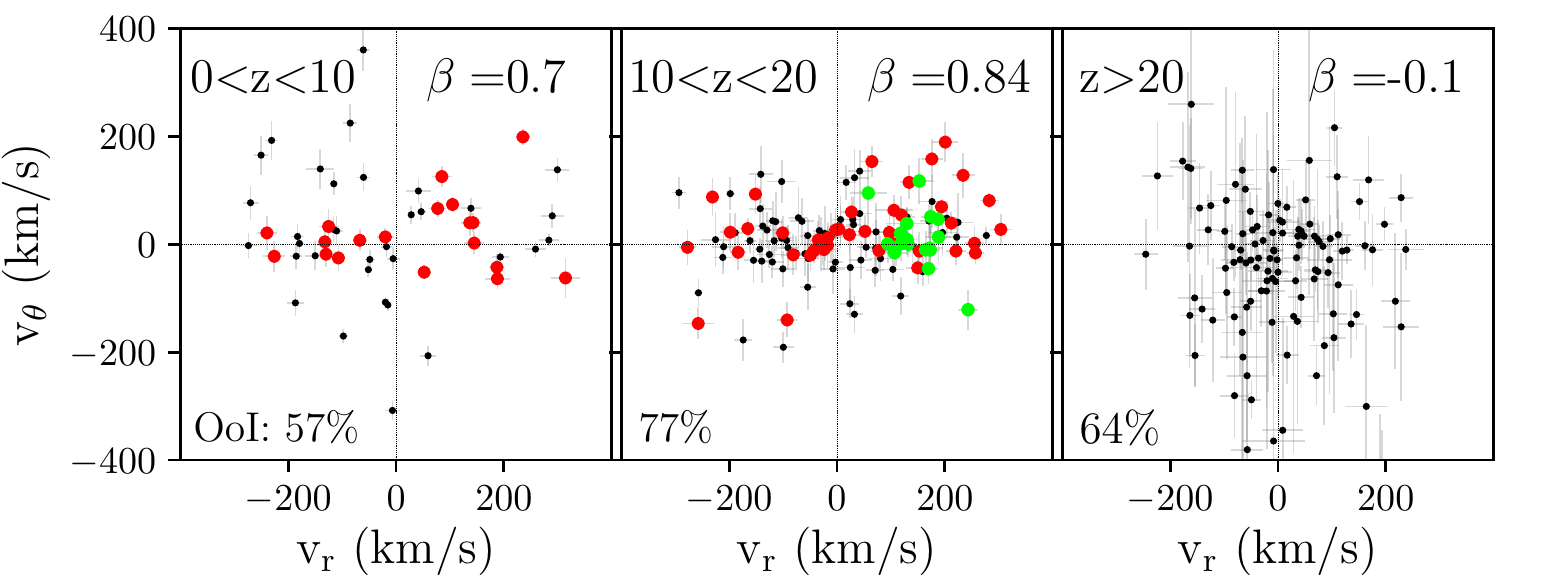}
\vspace{-0.45cm}
   \caption{Velocity distribution of the VOD RR Lyrae, in three
     height ranges above the Galactic plane, increasing from left to
     right ({\it cf.} the bottom left panel of Fig.
     \ref{fig:vel}). The fraction of Oosterhoff type I RR Lyrae within
     each distance range is reported on the bottom left of each panel
     and the orbital anisotropy $\beta$ in the top right. Note that in
     the middle panel, where $10<$z/kpc$<20$, the anisotropy
     parameter is $\beta=0.84\pm0.03$, close to the value reported for
     HAC (top panels of Fig.  \ref{fig:vel}); in the same field the
     large majority of RRL is of Oosterhoff type I.}
    \label{fig:VOD_vel}
\end{figure}

Using the virial theorem, \citet{actionhalo} concluded that the
radially anisotropic component of the stellar halo is also
significantly flattened vertically. To test this hypothesis, we split
the VOD sample into 3 groups according to their distance from the
Galactic plane and show the behaviour of the azimuthal $v_{\theta}$
and radial $v_{r}$ velocity distributions in
Figure~\ref{fig:VOD_vel}. Additionally, for each z slice we have
calculated the fraction of Oosterhoff type I (Oo I) RR Lyrae, using
equations 1 and 2 in \citet{Be2018} to explore the changes in the RRL
populations. In the 10$<$z/kpc$<$20 range, where the velocity anisotropy
is the highest ($\beta = 0.84 \pm 0.03$) approaching the value in the
HAC field, the Oo I type dominates with the 77 $\pm$ 13\% fraction. In the same
slice, 73\% of the stars belong to the more anisotropic (or `sausage'
looking) velocity ellipsoid. The slice closer to the galactic plane 
0$<$z/kpc$<$10 shows similar properties with anisotropy $\beta
= 0.7 \pm 0.1$ that is only moderately lower and the fraction of Oo I stars
that is broadly consistent with the above at 57 $\pm$ 21\%. Further from
the plane, at z$>$20 kpc, the velocity ellipsoid changes dramatically
to almost isotropic with $\beta = -0.1 \pm 0.2$ and Oo I type fraction
is 64 $\pm$ 13\%. We note however that in this particular z bin, the
$\beta$ value may be affected by the presence of the Sagittarius
stream. Group~2 and the stars sharing the same galactocentric distance
range with the HAC, are all located at $z<20$ kpc. Interestingly, the
HAC counterparts in the VOD (red points) are all clearly part of the
anisotropic component. To summarize, it is clear that the velocity
ellipsoid shape does change noticeably with vertical height. However,
given the limited range of Galactic $l$ and $b$ in the VOD sample, it
is not possible to conclude whether the structure is flattened or
simply ends around $z\sim20$ kpc.
\subsection{Orbital Properties}

\begin{figure}
	\includegraphics[scale=0.473]{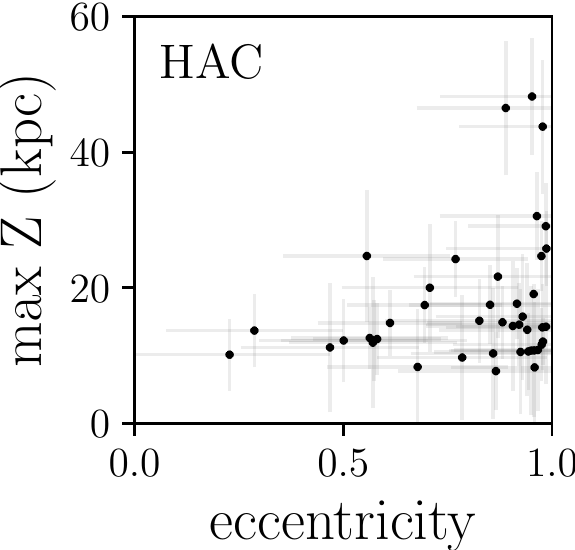}
    \includegraphics[scale=0.473]{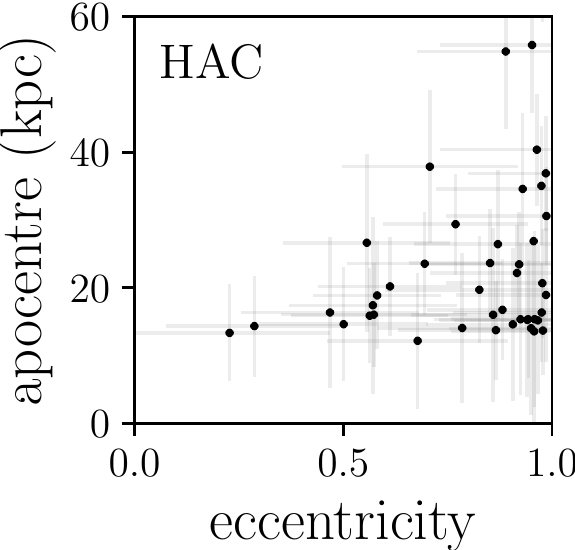} 
  \includegraphics[scale=0.473]{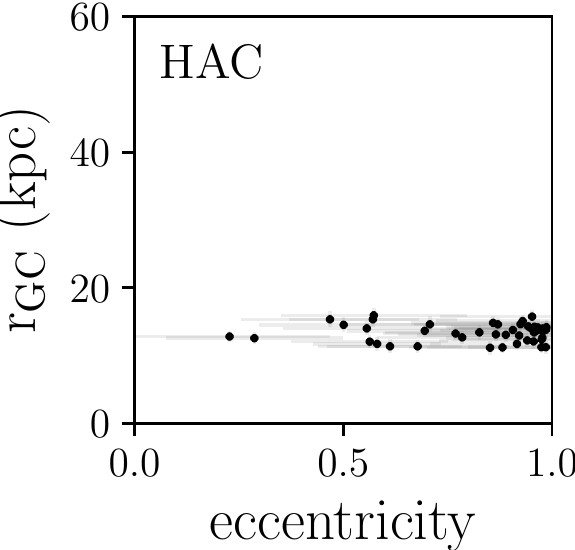} 
	\includegraphics[scale=0.473]{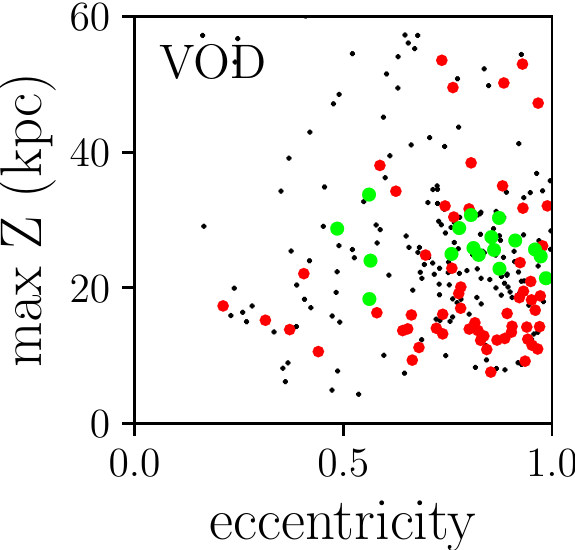}
    \includegraphics[scale=0.473]{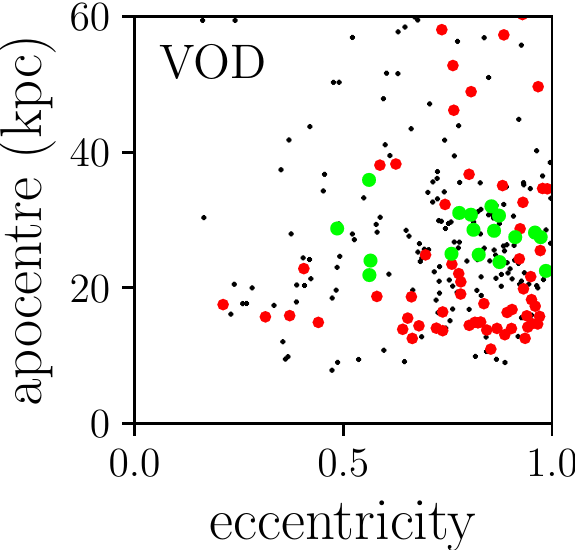} 
  \includegraphics[scale=0.473]{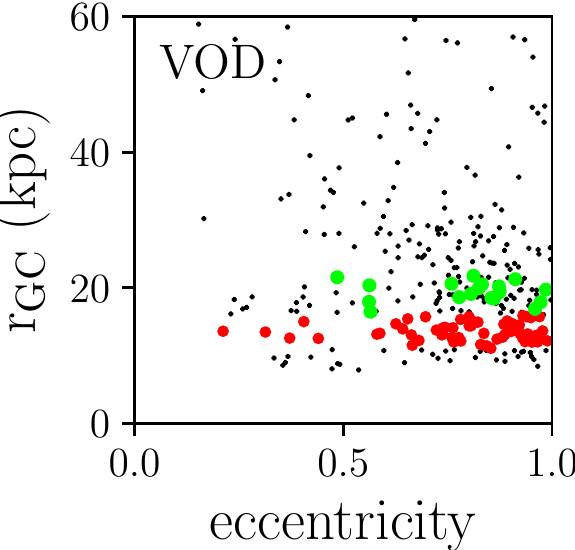} 
              \includegraphics[scale=0.473]{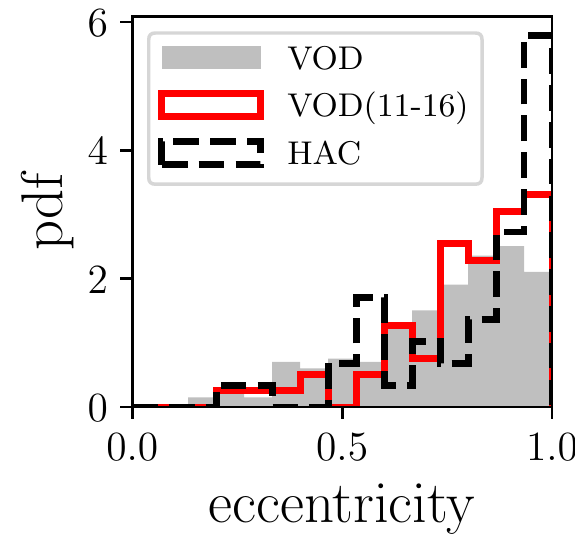} 
    \includegraphics[scale=0.473]{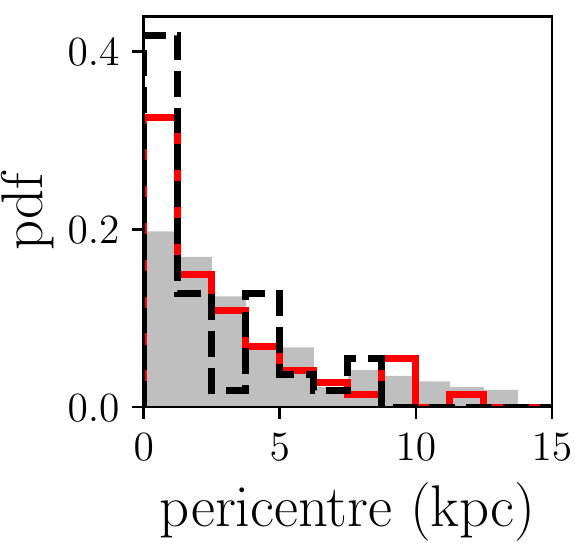} 
                          \includegraphics[scale=0.473]{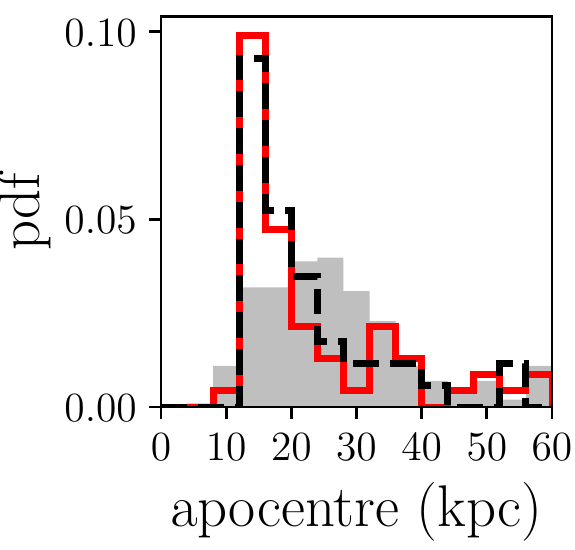} 
\vspace{-0.45cm}
  \caption{Properties of the RR Lyrae orbits in the HAC and VOD
    fields. In the top and middle rows we show the maximal height
    above the Galactic plane, apo-centre and Galactocentric radius as a
    function of eccentricity, for the HAC (top) and VOD (middle)
    stars. In the bottom row we show the probability distributions of
    the eccentricities, peri-centres and apo-centres for both fields,
    including the VOD(11-16) subsample (in red). The majority of HAC
    and VOD(11-16) stars are on highly eccentric orbits with small
    peri-centres (0-3 kpc from the GC) and apo-centres at 15-20 kpc.}
    \label{fig:orbits}   
    \end{figure}

We use the 6-D measurements described above to initialise the RR Lyrae
orbits in the HAC and VOD fields. The orbits are integrated using the
$\mathrm{galpy}$ package \citep{Bovy2015} in the recommended Galactic
potential model for the Milky Way, $\texttt{MWPotential2014}$, with
parameters given in Table 1 of \citet{Bovy2015}. The distribution of 
orbital
properties, i.e. the peri- and apo-centric radii, eccentricities, and
the maximal heights above the Galactic plane, are shown in
Figure~\ref{fig:orbits}. The uncertainties on orbital properties
(not shown for VOD to avoid cluttering the figure) are computed by 
integrating 500 orbits for each star with 
initial conditions sampled from observables according to
the associated uncertainties. The 1-D distributions of the
eccentricities, apo-centres and peri-centres are shown in the bottom
row of the Figure.  We remark that although the default Milky Way mass in
$\mathrm{galpy}$, $\mathrm{M_{vir}} = 0.8 \times 10^{12} M_{\odot}$,
is somewhat lower than suggested by recent measurements, we have
checked that the distributions of the orbital parameters
displayed in Figure~\ref{fig:orbits} are minimally affected if we
increase the Galaxy's mass.

The stars in the HAC field typically travel as high as $\sim$15 kpc
above the Galactic plane. The apo-centres bunch up around 15-20 kpc
from the Galactic center - similar to the behaviour of local Main
Sequence and BHB stars analysed in \citet{Deason2018pileup}, with a
tail to higher values. The distribution of the peri-centric distances
of the HAC stars peaks sharply around $\sim$1 kpc with most stars
having their peri-centers within 5 kpc from the Galactic center. This
naturally implies highly eccentric orbits (with eccentricity close to
1) for the vast majority of the HAC stars. Compared to the HAC, the
VOD stars explore broader range of apo-centres, with the distribution
of the peri-centres much less strongly peaked. As a result, the
typical eccentricity for a VOD star is around $\sim$0.8. Note however,
that a sample of VOD stars comprised of Group~2 (shown in green) and
stars that share current Galacto-centric distances with objects
in the HAC field appears to have orbital properties much closer to
that of HAC. In particular, once the stars with similar distances are
selected, the apo-centre and peri-center distance distributions in the
two fields look remarkably similar.

\begin{figure}
	        \includegraphics[scale=0.40352]{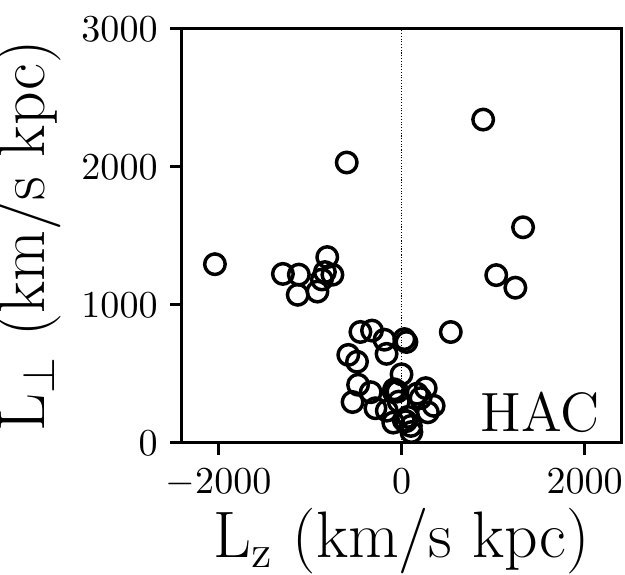}
	    \includegraphics[scale=0.40352]{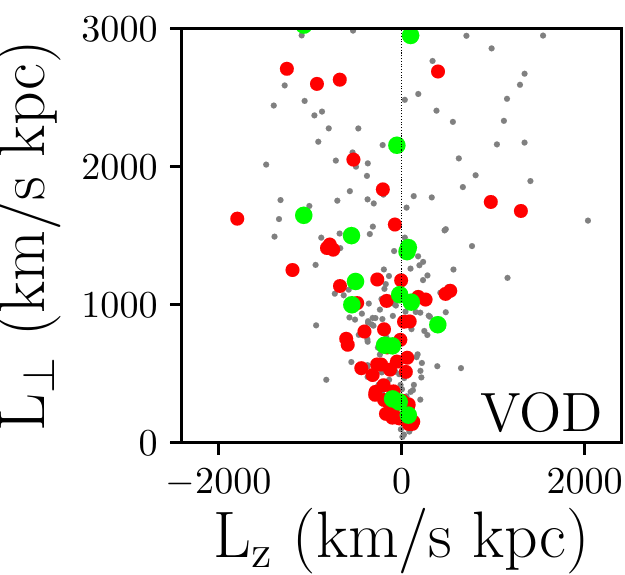}
        \includegraphics[scale=0.40352]{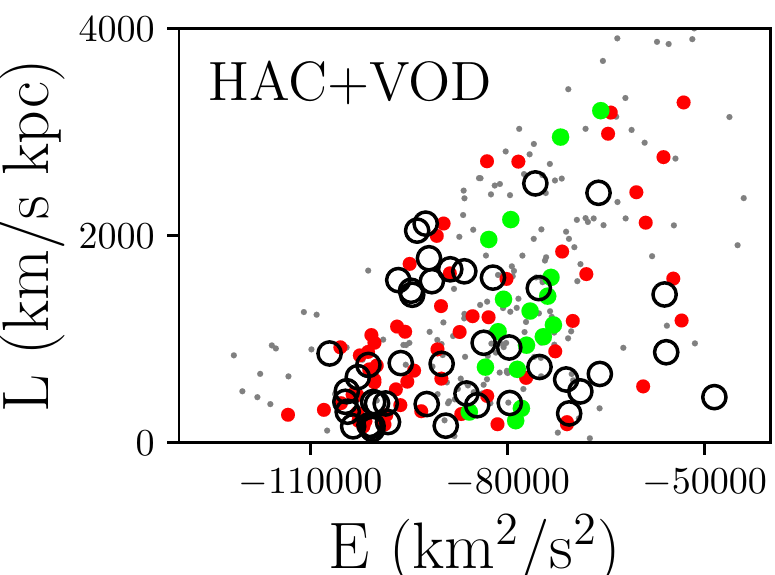}
		       	       	       	       \vspace{-0.46cm}
   \caption{Distribution of the angular momentum components,
     $\mathrm{L_{\perp}}$ and $\mathrm{L_{z}}$, for the HAC (left
     panel) and VOD (middle) RR Lyrae. Notice the prevalence of
     retrograde orbits for the HAC and VOD(11-16) stars. In the right
     panel, we show the total angular momentum as a function of energy
     for the HAC (black circles), VOD (gray dots) and the two VOD
     sub-samples, Vivas GroupGr 2 and VOD(11-16) marked in lime and red as
     in the previous figures. A large number of HAC and VOD(11-16) RR
     Lyrae are concentrated at high energy E$\sim$-100000
     km$^{2}$/s$^{2}$ but low angular momentum L$\sim$0-500 km/s kpc.}
    \label{fig:energy}
\end{figure}

The similarity of orbital properties of the HAC and VOD stars is
emphasized in Figure~\ref{fig:energy}. Here, we first show the
distribution of the components of the angular momentum $L_{\rm z}$ and
$L_{\perp}$ for HAC (VOD) stars in the left (middle) panels. The stars in the 
VOD subset with distances matching HAC stars $11\mathrm{<r_{GC}}/$kpc$<16$ are shown in red. The
right panel presents the behaviour of the total angular momentum as a
function of energy. The bulk of the stellar debris in both fields are
on high energy, low angular momentum orbits. As evident from the
Figure, while stars in both HAC and VOD occupy highly radial orbits,
both prograde and retrograde objects exist, with a slight prevalence
of the retrograde ones \citep[see
  also][]{actionhalo,shards,Helmi2018}.

\begin{figure}
	     \includegraphics[scale=0.302]{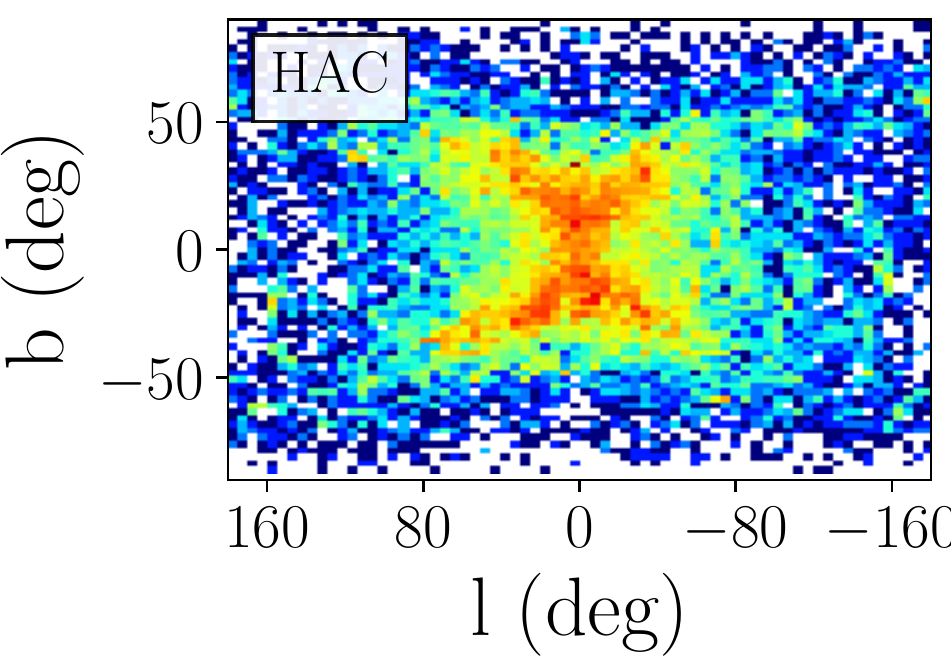}
             \includegraphics[scale=0.302]{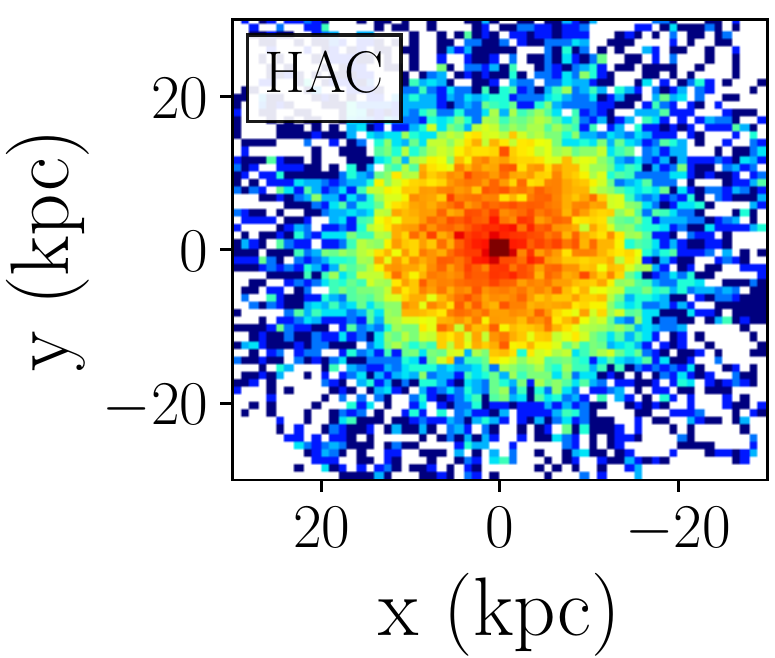}
             \includegraphics[scale=0.302]{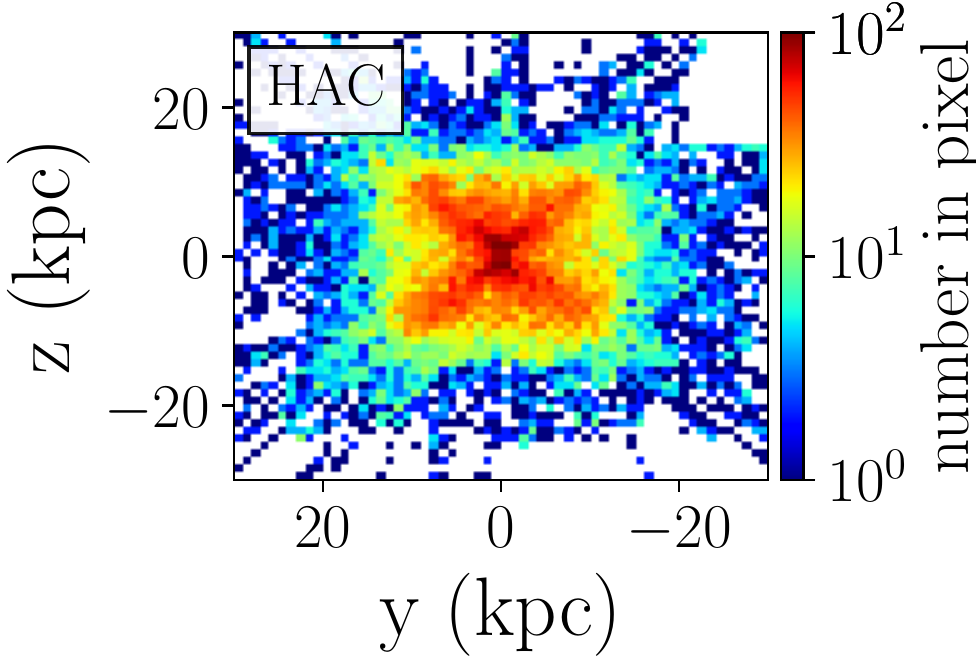}
             \\ \includegraphics[scale=0.302]{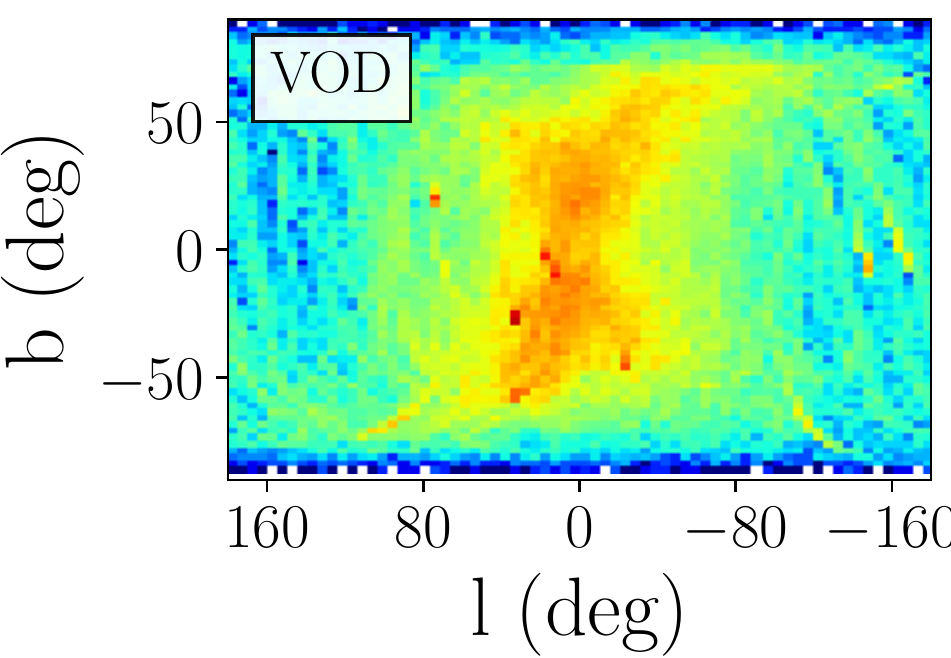}
             \includegraphics[scale=0.302]{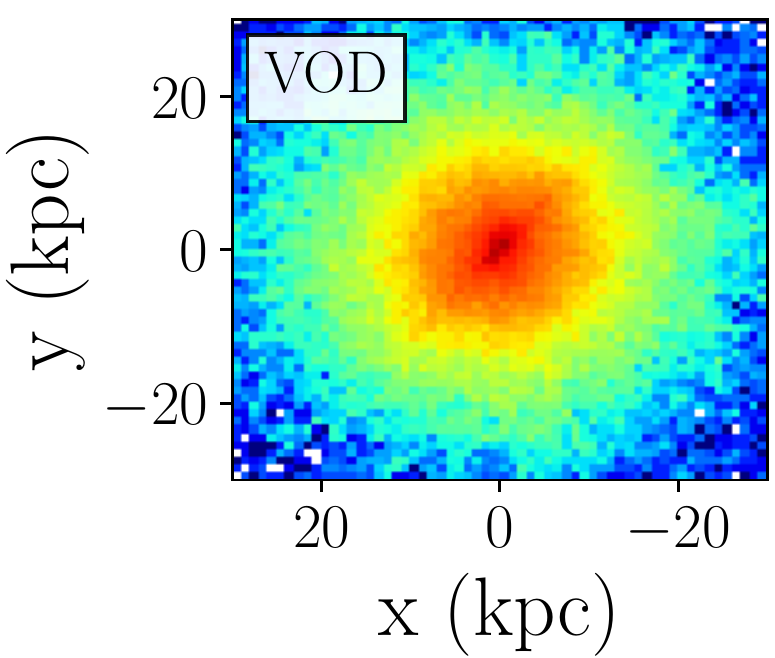}
             \includegraphics[scale=0.302]{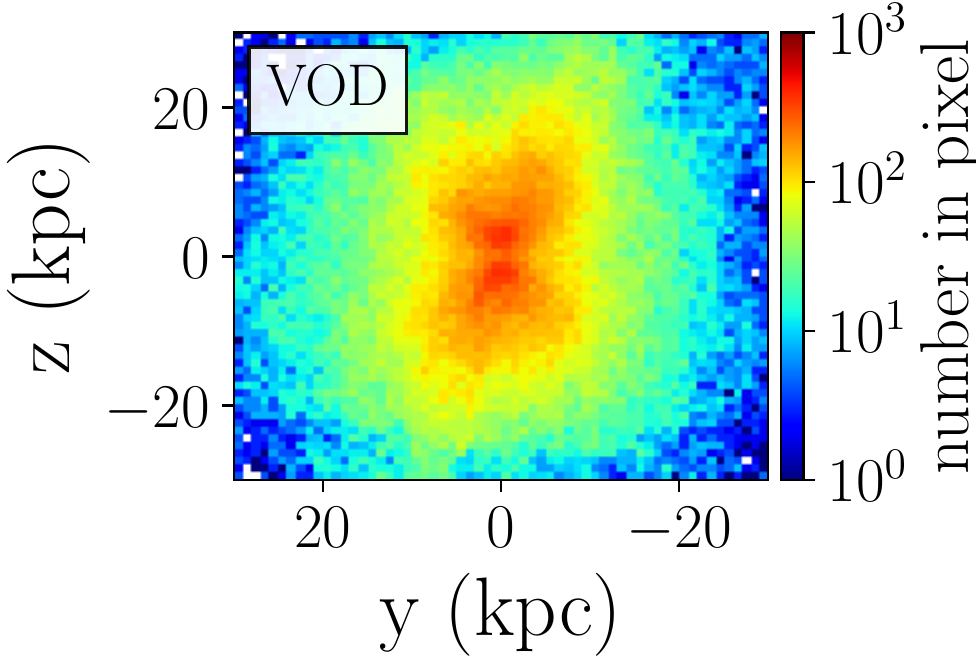}\\ \includegraphics[scale=0.302]{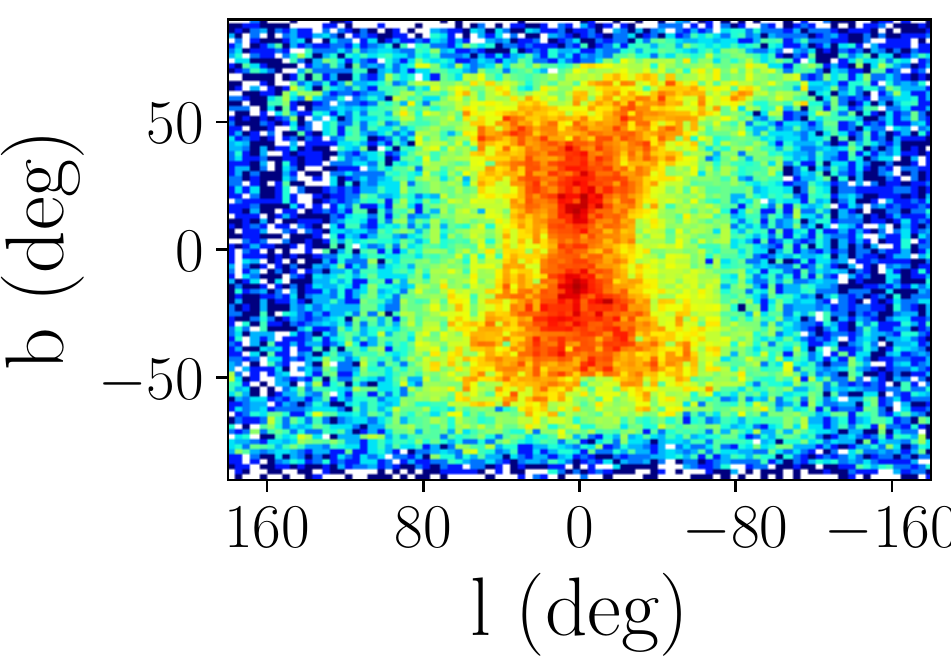}
             \includegraphics[scale=0.302]{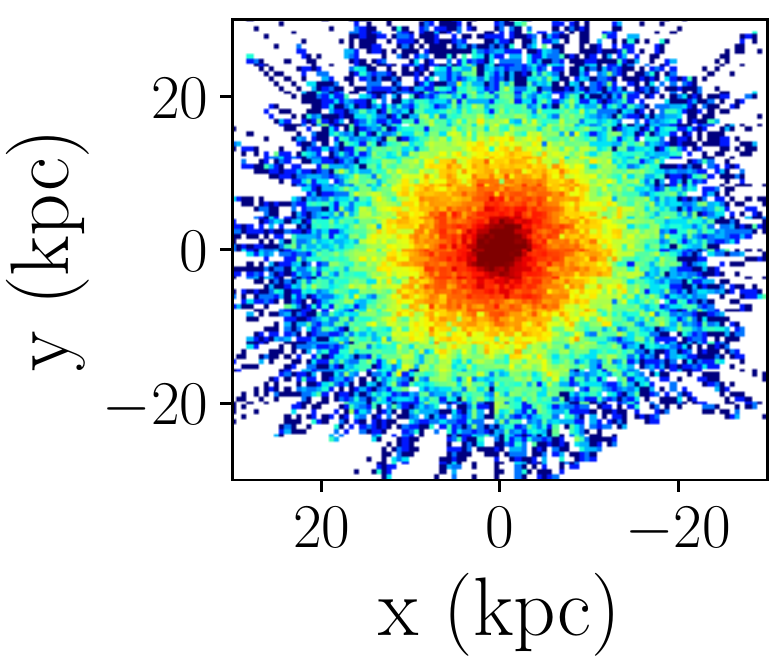}
             \includegraphics[scale=0.302]{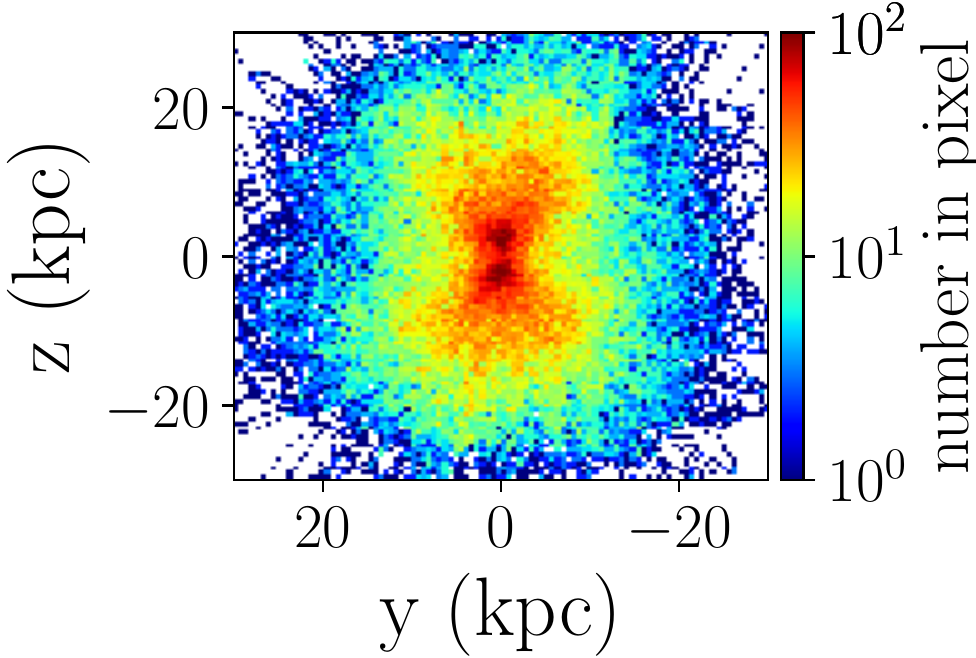}
\vspace{-0.45cm}
   \caption{Density maps of the RR Lyrae positions along their
     individual orbits over the past 8 Gyrs, in Galactic coordinates
     on the sky (left panels) and in the x-y (middle) and y-z (right) Galactic
     planes.  The top and middle rows show the integrated orbits for
     all the HAC and VOD stars while on the bottom row we show only
     the orbits for the VOD stars in the `sausage' component (marked
     in blue in Fig. \ref{fig:vel}). Note that the observed orbital
     morphology of the stellar debris agrees well with the simulated
     merger event shown in Figure 7 of \citet{Simion2018}.}
    \label{fig:backorbits}
\end{figure}

Figure~\ref{fig:backorbits} provides an alternative view of the
orbital properties of the HAC and VOD stars. Here we have used the
stellar 6-D phase-space measurements as initial conditions and
integrated the orbits back in time for 8 Gyrs. The Figure shows the
density of the test particle positions along each orbit for all orbits
across the entire temporal range of the integration. The HAC stars
occupy a slightly flattened in the vertical direction and cross-like in
the y-z plane and its projection on the sky structure
with a density peak at the Galactic centre. Note that some of the
appearance of the distribution of debris could be caused by limited
footprint and therefore initial conditions of stars with available radial velocity
measurements. The VOD stars typically move through very similar
regions of the Galaxy (especially the sausage sub-sample shown in the
bottom row). While, overall, the match between the stellar debris
distributions in Figure~\ref{fig:backorbits} is striking, the HAC and
VOD stars studied here do not have identical orbital properties. For
example, the VOD stars travel further above the Galactic disc, as
demonstrated by much rounder, or perhaps even vertically stretched, y-z
distribution (right panels).
Notwithstanding possible selection effects present in the data, the
stellar density distribution shown in Figure~\ref{fig:backorbits}
looks staggeringly similar to the debris distribution of the simulated
accretion event presented in Figure 7 of \citet{Simion2018}. This
particular merger happened some 11 Gyr ago according to the suite of
numerical stellar halo formation models of \citet{Bu05}. Giving the
conspicuous similarity in the appearance of the orbital density
distributions of the HAC and the VOD stars and the simulated merger
example, we conclude that the Hercules-Aquila Cloud and the Virgo
Overdensity are both parts of one ancient massive head-on collision.
\section{Conclusions}
Using a sample of $\sim$350 RR Lyrae with full 6-D phase space
information, we have studied the orbital properties of the
Hercules-Aquila and Virgo Clouds. Both Clouds appear dominated by
stars on highly eccentric orbits. Assuming that the kinematics of each
structure is well described by a single Gaussian, the orbital
anisotropy of the HAC is $\beta=0.91$ and for the VOD,
$\beta=0.74$. Note, however that the original criteria applied to the
CRTS RR Lyrae dataset to select targets for spectroscopic follow-up
differ drastically between the HAC and the VOD datasets. The HAC
sample covers a very limited region in the of $l,b, D$ space, while
the VOD dataset spans a wide range of longitudes, latitudes and
heliocentric distances. It is therefore likely that the VOD dataset
contains a mixture of several halo sub-structures \citep[see][for a
  detailed discussion]{Vivas2016}. For the entirety of the analysis
described here, we made sure to cull the probable Sgr stream
members. Additionally, we explore how the VOD's make-up changes with
Galactic height and demonstrate that for $|z|<20$ kpc, the VOD orbital
anisotropy is $\beta\sim 0.8$, while above this threshold, it quickly
changes to $\beta\sim0$. We conclude therefore, that an assumption of
a single Gaussian for the entire VOD sample is likely not
appropriate. Modeling the kinematics of the VOD stars with a mixture
of 2 multivariate Gaussians, we show that the VOD can be separated into 
two components: one with two thirds of the stars and $\beta=0.96$, and the 
 other one with $\beta=0.44$, in good agreement 
with the local measurement presented
in \citet{Belokurov2018}.

As revealed by Gaia, the two structures are composed of stars on
nearly radial orbits, with peaks in the eccentricity distribution at
0.95 (0.8) for the HAC (VOD). The distributions of the peri-centric
and apo-centric distances also match: the stars in the Clouds turn
around at $1-2$ and $15-25$ kpc. Not only the HAC and the VOD look
alike kinematically, their orbital composition is in perfect agreement
with the stellar halo properties as analysed locally by
\citet{Belokurov2018} and globally (out to 40 kpc) by
\citet{Deason2018pileup}. As these authors demonstrate, the inner halo
is dominated by metal-rich debris from an old and massive accretion
event. In particular, \citet{Belokurov2018} use Cosmological
simulations of the Milky Way halo formation, to bracket the time of
the merger - between 8 and 11 Gyr ago - and its mass, which they show
to be in excess of $10^{10} M_{\odot}$. The tell-tale sign of this
dramatic head-on collision is the particular shape of the
corresponding stellar velocity ellipsoid, which is stretched so much
in the radial direction (compared to the tangential ones), that it
resembles a sausage. An alternative view of this merger can be found
in \citet{actionhalo}, where the local stellar halo is mapped out in
the action space. Here, the metal-rich stars are shown to have an
extended radial action distribution in addition to a prominent spray
of material on retrograde orbits. The high mass of the progenitor is
evidenced not only by the metallicity distribution of its likely
members or the numerical simulations of halo formation, but also by a
sizeable number of Globular Clusters that could be attributed to the
same event \citep[see][]{sausagegc,Kruijssen2018}.

Our interpretation of the nature of the HAC and VOD is in broad
agreement with the earlier studies of \citet{Jo2012} and \citet{Ca12}
whereby each group of authors have singled out a progenitor on a high
eccentricity orbit as a culprit for the production of HAC and VOD
respectively. However, for the first time, we connect both of the
discussed debris Clouds to a single event with yet higher eccentricity
and yet larger mass. Additionally, rather than being a recent
accretion, the ``sausage'' merger likely happened around the epoch of
the Galactic disc formation, i.e. between 8 and 11 Gyrs ago
\citep[see][]{Belokurov2018,Helmi2018}. As seen by Gaia, the HAC and
the VOD represent two poorly-mixed portions of the large amount of
tidal material dumped onto the Milky Way in that event. While the two
Clouds look remarkably similar in the space of the integrals of
motion, their orbital properties are not exactly identical. Some of
the mismatch could perhaps be attributed to the selection
effects. However it is not impossible that the differences we are
seeing are related to the details of the merger event itself, e.g. the
effects of variable stripping time or the orbital evolution of the
progenitor under the action of dynamical friction. Future models of
the Clouds' orbital properties \citep[along the lines of the ideas
  laid out in e.g.][]{Jo2012,Sanderson2013} will inform our
understanding of this dramatic head-on collision that re-shaped the
Galaxy.

\section*{Acknowledgements}

We wish to thank Alis Deason, Kathryn Johnston and Wyn Evans for
insightful comments that helped to improve the quality of the
manuscript.  The research leading to these results has received
funding from the European Research Council under the European Union's
Seventh Framework Programme (FP/2007-2013), ERC Grant Agreement
n. 308024 and PIFI Grant n. 2018PM0050. ITS and SEK thank the Center for Computational Astrophysics
for hospitality.



\bibliographystyle{mn2e}
\bibliography{bibl}  
\bsp	
\label{lastpage}
\end{document}